**Diffusion behavior of Li ions in crystalline and amorphous Li-Zr-O and Li-Nb-O phases**

Daniel Mutter[1], Diego A. Pantano[2], Christian Elsässer[1], Daniel F. Urban[1]

[1]Fraunhofer IWM, Wöhlerstraße 11, 79108 Freiburg, Germany

[2]TotalEnergies OneTech, 2 place Jean Millier, 92400 Courbevoie, France

Abstract

Li containing transition metal oxides are known as good ionic conductors. Performing classical molecular dynamics simulations, the diffusion behavior of Li ions is investigated in crystalline and amorphous phases with the stoichiometries $Li_2ZrO_3$ and $LiNbO_3$. We first demonstrate the stability of the crystal structures for the used interatomic potential model and then analyze the amorphous phases, which result from melt-and-quench simulations, in terms of radial distribution functions. Diffusivities of Li ions in those systems are obtained from a statistical Arrhenius analysis of mean square displacement curves at different temperatures. The crystalline phase of $Li_2ZrO_3$ exhibits two well-defined migration mechanisms: vacancy-mediated migration is dominant below and a site exchange of Li ions above a crossover region between about 1700 and 1800 K. The latter mechanism also prevails in the amorphous phases of $Li_2ZrO_3$ with a strongly reduced activation energy, which is due to a smaller equilibrium separation of Li ions as in the crystal structure. This migration mechanism is found in amorphous $LiNbO_3$, too.

## 1. Introduction

Surface coatings are widely applied on the active cathode particles of Li-ion batteries (LiBs) to improve their electrochemical performance in loading, unloading, and repeated cycling [1]. Numerous studies report on crystalline $Li_2ZrO_3$ (c-LZO) [2]–[5] and $LiNbO_3$ (c-LNO) compounds [6], [7] deposited as thin protective layers on NCM- ($Li[Ni_xCo_yMn_z]O_2$) or LCO- ($LiCoO_2$) type cathodes. The layers were shown to diminish detrimental side reactions between the electrodes and the electrolyte and to increase the structural and thermal stability of the cathodes, leading to reduced capacity fading compared to LiBs with uncoated NCM- or LCO-type materials. If the structure of the coating layers is amorphous (a-LZO, a-LNO), the systems benefit from an improved interface compatibility, and, accordingly, reduced interfacial stress [8]–[10], which is particularly relevant in combination with solid electrolytes (SEs) in all-solid-state batteries [11]–[14]. The protective layers on the cathode particles act as thin SE phases. To ensure a good electrochemical performance, these phases must therefore exhibit a sufficiently high Li-ion conductivity [15], [16].

Monoclinic c-LZO is long known as an ionic conductor [17]. Ferreira et al. [18] determined an energy barrier of 0.75 eV for vacancy-mediated Li migration from density functional theory (DFT) calculations. Nuclear magnetic resonance (NMR) studies revealed a substantial concentration of vacancies and an uneven distribution between the two inequivalent crystallographic Li sites in c-LZO [19], which was assumed by Sherstobitov et al. as the origin for the observed high Li ion conductivity at increasing temperatures [20]. Even for vacancy-mediated diffusion, rather high Li ion migration energies of about 1.3 eV were reported for c-LNO, both experimentally [21] and from DFT calculations [22], which would make this material unsuitable as efficient ionic conductor. In amorphous phases (a-LNO) however, conductivities can reach values which are higher by many orders of magnitude than those of the crystalline phases, corresponding to energy barriers in the range of only 0.5 – 0.8 eV [23]–[26].

For amorphous materials, the atomic self-diffusion, i.e., the mobility of elements constituting a compound (in contrast to e.g. impurities), is described in the literature by different mechanisms. The concept of vacancies or interstitials as diffusion-mediating point defects, which are well defined in crystal structures, cannot be adopted to amorphous phases due to the lack of a reference structure [27]. Instead, there can be excess free volume (i.e., locally a lower atom density than in the corresponding crystalline phase) facilitating the movement of atoms, typically in a cooperative process involving more than two atoms [28], which leads to higher conductivities than in the case of individual jumps of atoms or ions [29], [30]. For example, in amorphous Si (a-Si) the atoms move by exchanges of nearest neighbors via collective rearrangement, breaking, and switching of bonds [31]–[33]. In metallic glasses, such as a-Cu-Zr, molecular dynamics (MD) simulations revealed a chain process of at least 10 atoms, each being displaced by only a fraction of the nearest neighbor distance, resulting in a net diffusion [34], [35]. The higher the temperature, the more atoms are involved in such a process [36]. Vogel reports Li ion migration in a-$LiPO_3$ as correlated jumps across a network of nearly unmodified sites in the rigid $PO_3$-glass matrix [37], [38]. To the best of the authors' knowledge, there are no reports so far about MD simulations elucidating the migration processes in a-LZO and a-LNO.

For such collective, multi-atom processes occurring in amorphous phases in not well-defined structural environments , a computational analysis via static calculations of energy barriers for individual atomic migration events, e.g., using the nudged elastic band method [39], is cumbersome, especially since the mechanisms are typically not known beforehand. To obtain information about diffusion processes and activation energies in amorphous phases, it is more convenient to do MD simulations, where the migration processes can be directly observed, and an averaged activation energy and diffusivity can be extracted via the mean square displacement of atomic species. Especially at low temperatures, this can however require long simulation times in the range of nano- to microseconds to detect a sufficiently high number of jumps for the analysis. Also, an average over many configurations is desirable to ensure good statistics, and, in the case of amorphous phases, large simulation cells should be used to avoid an artificial long-range order resulting from periodic boundary conditions. These requirements can hardly be fulfilled by first-principles MD simulations based on DFT. In this work, we instead applied MD simulations using a Morse potential derived from the bond valence (BV) theory [40]. The BV method is widely used for static analysis and visualization of migration paths in crystalline ionic conductors [22], [41], but its implications for dynamic migration processes, both in crystalline and in amorphous structures, are scarcely addressed [40], [42].

The paper is organized as follows: after a description of the employed computational methods and the obtained crystalline and amorphous structures in Section 2, the results of the Li-ion-diffusion calculations are reported and discussed for c-LZO in Section 3.1 and for a-LZO and a-LNO in Section 3.2.

## 2. Computational Method and Structure Models

### 2.1. Molecular Dynamics Simulations

The MD simulations were performed with LAMMPS [43], and interatomic interactions were modelled with a Morse potential. This potential is derived from the BV method, which describes chemical bonds of cations and anions by the postulate that the oxidation state of an ion equals the sum of bond valences between this ion and its nearest neighbors of opposite charge [44]. An energy scale can be linked to the deviation of the sum of distance-dependent bond valences from the ideal oxidation state, describing the attractive part of the interionic potential. The repulsion of equally charged ions is considered by a screened electrostatic interaction. In Ref. [40], Adams and Rao describe this method in detail. Additionally, they provide parameters for the interaction of oxygen with an extensive list of cations, from which we adopted the values for the ionic pairs $Li^{+}$-$O^{2-}$, $Zr^{4+}$-$O^{2-}$, and $Nb^{5+}$-$O^{2-}$ for the simulations performed in this work. MD simulations were carried out with a timestep of 1 fs in NVT and NPT ensembles at various temperatures, and, in the case of NPT, at a pressure of 0 Pa. Periodic boundary conditions were applied to the simulation cells in all three spatial directions. We recorded the mean square displacements (MSDs) of the different elemental species during the simulation which are linearly related to their diffusivity [45].

## 2.2. Structures

c-LZO with the stoichiometry $Li_2ZrO_3$ was set up in the monoclinic structure with space group $C2/c$ (s.g. #15), at first in the conventional cell containing 4 formula units (f.u.) (Figure 1, top right), with lattice parameters and relative positions of the atoms as obtained from experimental neutron-diffraction data [46]. Using a fully flexible simulation cell, the structure was heated up to 500 K in the NPT ensemble (with a rate of 5000 K/ns), equilibrated at this temperature for 0.1 ns, and cooled down again to 1 K with the same rate, followed by a static structural relaxation. Table 1 lists the resulting lattice parameters and atomic coordinates in comparison to the experimental data. The monoclinic structure of c-LZO as well as the atomic positions are very well preserved in the heating-and-cooling simulation, however, the potential predicts that the edge lengths of the cell and the monoclinic angle are lower in the range of 1.6 – 7.0 %. Since the purpose of this work is more the analysis of migration mechanisms than an accurate quantitative prediction of diffusivities, we considered the Morse potential to be suitable despite this apparent structural deviation. Note also that the potential parameters were not explicitly optimized to reproduce the monoclinic structure of c-LZO but derived using sets of reference structures of Li oxides and Zr oxides separately. Hence, it is remarkable that the underlying simple chemical notion of bond valences and electrostatic interactions leads to a potential for the ternary compound, which predicts its complicated monoclinic structure reasonably well as a local energy minimum. This demonstrates the transferability of the potential, which is why we adopted it also for constructing the amorphous phase.

*Table 1. Lattice parameters and relative atomic coordinates of c-LZO resulting from MD simulations with the Morse potential used in this work and from experiments (Ref.* [46]*). The specifiers in parentheses next to the elements denote their Wyckoff positions, and the three numbers given are the relative x, y, and z coordinates of a representative site in the monoclinic simulation cell.*

| | Morse potential | Experiment |
|---|---|---|
| $a$ (Å) | 5.040 | 5.422 |
| $b$ (Å) | 8.875 | 9.022 |
| $c$ (Å) | 5.227 | 5.419 |
| $\alpha$ (°) | 90.0 | 90.0 |
| $\beta$ (°) | 109.3 | 112.7 |
| $\gamma$ (°) | 90.0 | 90.0 |
| Li1 (4*e*) | 0.000, 0.418, 0.250 | 0.000, 0.423, 0.250 |
| Li2 (4*e*) | 0.000, 0.744, 0.250 | 0.000, 0.742, 0.250 |
| Zr (4*e*) | 0.000, 0.091, 0.250 | 0.000, 0.092, 0.250 |
| O1 (4*d*) | 0.250, 0.250, 0.500 | 0.250, 0.250, 0.500 |
| O2 (8*f*) | 0.260, 0.580, 0.507 | 0.272, 0.575, 0.486 |

c-LNO ($LiNbO_3$) crystallizes in a trigonal structure with space group $R3c$ (s.g. #161). Table 2 lists the experimentally derived lattice parameters of the conventional cell, which contains 6 f.u. (Figure 2, top right), and the relative atomic positions [47], together with the values obtained from a heating-and-cooling equilibration and optimization of the cell using the Morse potential similar as described above for c-LZO. The trigonal symmetry of the structure is maintained, and only the positions of the oxygen atoms deviate slightly. As in the case of c-LZO, the edge lengths of the cell are predicted by the Morse potential to be lower as in the experiment, however only by 0.7 for $a$ and by 3.8 % for $c$.

*Table 2. Lattice parameters and relative atomic coordinates of c-LNO resulting from MD simulations with the Morse potential used in this work and from experiments (Ref.* [47]*). The specifiers in parentheses next to the elements denote their Wyckoff positions, and the three numbers given are the relative x, y, and z coordinates of a representative site in the trigonal simulation cell.*

| | Morse potential | Experiment |
|---|---|---|
| $a$ (Å) | 5.114 | 5.149 |
| $b$ (Å) | 5.113 | 5.149 |
| $c$ (Å) | 13.338 | 13.863 |
| $\alpha$ (°) | 90.0 | 90.0 |
| $\beta$ (°) | 90.0 | 90.0 |
| $\gamma$ (°) | 120.0 | 120.0 |
| Li (6$a$) | 0.000, 0.000, 0.285 | 0.000, 0.000, 0.283 |
| Nb (6$a$) | 0.000, 0.000, 0.003 | 0.000, 0.000, 0.000 |
| O (18$b$) | 0.059, 0.354, 0.070 | 0.049, 0.345, 0.065 |

Typically, amorphization is reached in MD simulations by following a melt-quench protocol [48]–[50], where first a crystal structure is heated above its melting point, followed by an equilibration of the liquid phase and subsequent cooling with a rate high enough to prevent recrystallization. Such amorphous phases exhibit an artificial long-range order due to the finite size and periodic boundary conditions of the simulation box. However, starting the melt-quench approach from a crystal phase in a non-cubic cell in general imposes an anisotropy on the final amorphous systems in the form of different edge lengths and angles of the simulation box, even in case of performing NPT-MD with fully flexible cell size and shape. Amorphous phases exhibit spatial long-range isotropy, which can best be ensured in the simulations by using cells of cubic shape and allowing only isotropic cell variations.

Since in general, one obviously cannot build cubic cells for non-cubic crystal structures, in this work, we instead set up cubic boxes and filled them with random distributions of atoms in the desired stoichiometries, followed by isotropic NPT equilibration at temperatures corresponding to the liquid phase [51]. The initial box sizes were chosen such that they lead to densities a bit below those of the liquid phases known beforehand from melting the crystal structures. The averaged energies and volumes per f.u. of the equilibrated liquid phases at a specific temperature were identical when either a melting of a crystal phase was performed, or an initial random atomic configuration was used. The latter approach has the advantage of an arbitrary choice of the number of formula units, since it is not limited to integer multiples of the primitive cell to build a supercell whose shape is as cubic as possible. Following this procedure, we set up cubic boxes with 1800 atoms for constructing a-LZO (300 f.u.) and of 2000 atoms for a-LNO (400 f.u.). Ensembles of 10 configurations each were considered. After equilibrating the liquid phases at temperatures of 3000 K, they were cooled down with rates of $-1000$ K/ns (LZO) and $-100$ K/ns (LNO). Cooling LZO with $-100$ K/ns resulted in recrystallization for eight out of the ten systems, which is why a faster rate was chosen in this case. Such a recrystallization did not happen in any of the ten LNO systems.

To analyze the amorphous phases, we calculated both total and pairwise radial distribution functions (RDFs) and averaged them over the 10 samples. The results are shown in Figure 1 for LZO and in Figure 2 for LNO, in both cases together with the RDFs of the crystalline phases for comparison. As expected, the long-range distance relationships between the atoms are considerably smoothed out in the amorphous phases, but short-range order is apparent for Li-O and Zr/Nb-O nearest-neighbor configurations with distances close to the values of the crystal phases. In both materials, the Li-Li peaks considerably shift to smaller values upon amorphization. In the LNO system, this effect is also observed for the Nb-Nb peak, although not as pronounced as for Li-Li, whereas the Zr-Zr peaks in a-LZO and c-LZO are at similar positions.

In c-LZO, Zr ions are octahedrally, i.e., 6-fold-, coordinated by O, corresponding to the first Li-O peak in the pairwise RDF, located between 2.0 and 2.2 Å. The Li-O peak in a-LZO extends up to about 2.4 Å. Defining the nearest-neighbor environment by this value, one finds about 50 % of the Zr ions to be 7-fold- and 50 % to be 8-fold coordinated by O. This resembles the bonding situations in different stable and metastable variants of pure $ZrO_2$, as for example in the cubic structure (s.g. $Fm\overline{3}m$ [52]) the coordination of Zr by O is 8-fold, while it is 7-fold in the monoclinic (baddeleyite, s.g. $P2_1/c$ [53]) and orthorhombic (s.g. $Pca2_1$ [54]) structures. In the latter two, the corresponding coordination polyhedra are edge- and corner-sharing, which is also observed for the a-LZO phases, as visualized in Figure 1. Li ions are distributed in this network such that on average smaller Li-Li distances occur than in c-LZO. Even though a-LZO has been used repeatedly as protective layer on cathode materials (see Section 1), there is, to the best of our knowledge, no report on measured or computed RDFs of a-LZO in the literature, nor did we find any detailed structural characterization of this amorphous compound with information about ionic coordination.

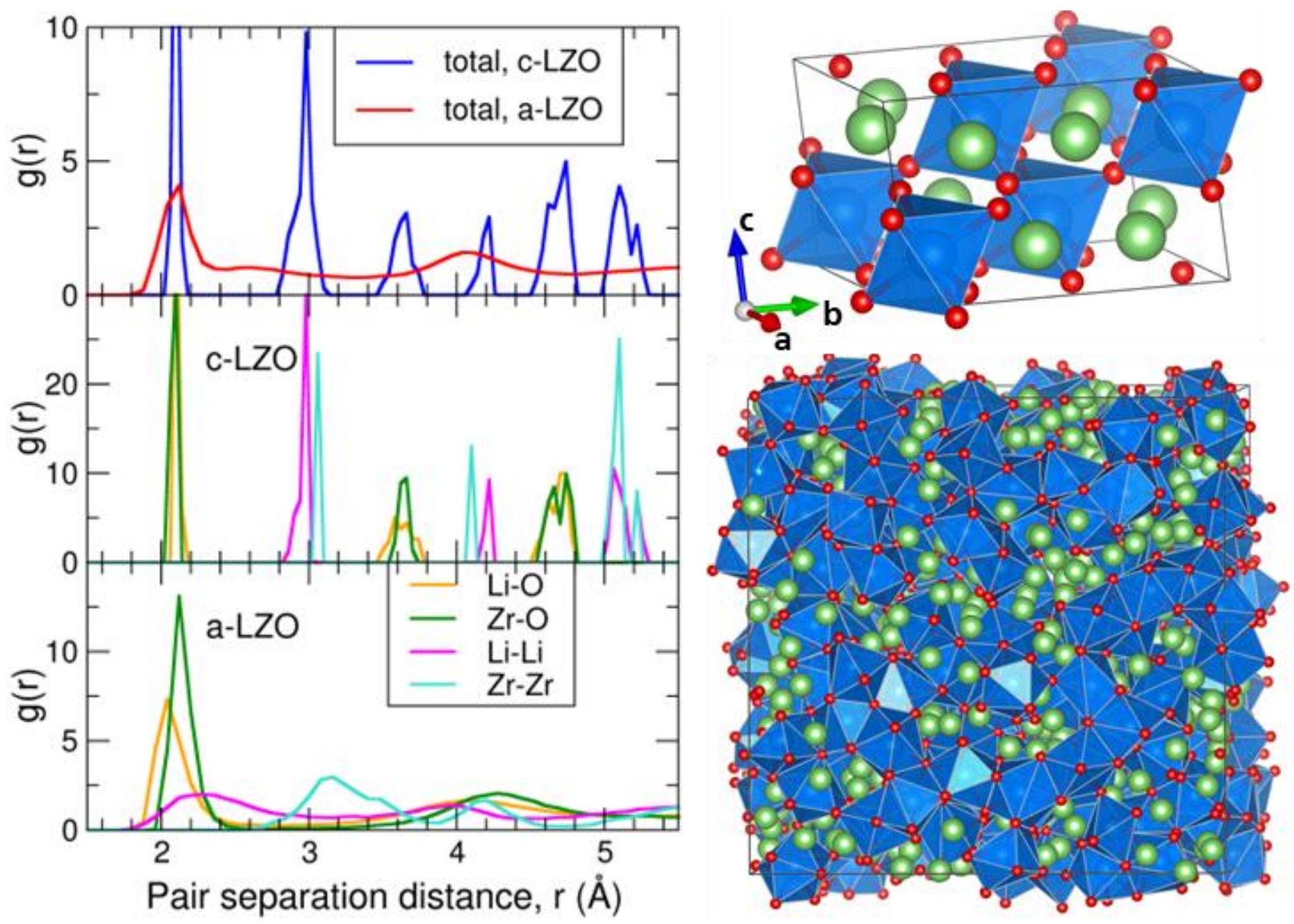

*Figure 1. Left: Total (top) and pairwise (center and bottom) RDFs of c-LZO and of a-LZO (average over 10 samples), in both cases for a bin size of 0.04 Å. Right: conventional monoclinic cell of c-LZO (top) and cubic box of a-LZO (bottom) with Li atoms in green, O atoms in red, and O-coordinated polyhedra around Zr atoms in blue.*

The total and pairwise RDFs of a-LNO predicted by the model are very similar to those of a-LZO. In a-LNO, there are also edge- and corner-shared $NbO_n$ polyhedra which build up the structure (see Figure 2). However, an analysis of the nearest-neighbor environment of the Nb ions (as for a-LZO defined by a radius of 2.4 Å around them) yields about 60 % of 8-fold-, and each 20 % of both, 7- and 9-fold-O-coordinations. While, as for a-LZO, we are not aware of measured or differently computed RDFs of a-LNO, which would allow a direct comparison, a few reports state that this phase contains 6-fold coordinated Nb ions forming an irregular network with Li located in between [24], [55]. Such a Nb-O bonding environment is at variance to the configuration predicted by our model, but 7- and 8-fold-coordinated Nb was observed as building blocks of amorphous $Nb_2O_5$ [56], where Nb is in the charge state +5 as in LNO. Due to the lack of more detailed structural information on a-LNO, and since the Li ions are distributed irregularly in between the polyhedra as it is reported, we considered the obtained structure to be acceptable for the further analysis of Li-ion conductivity.

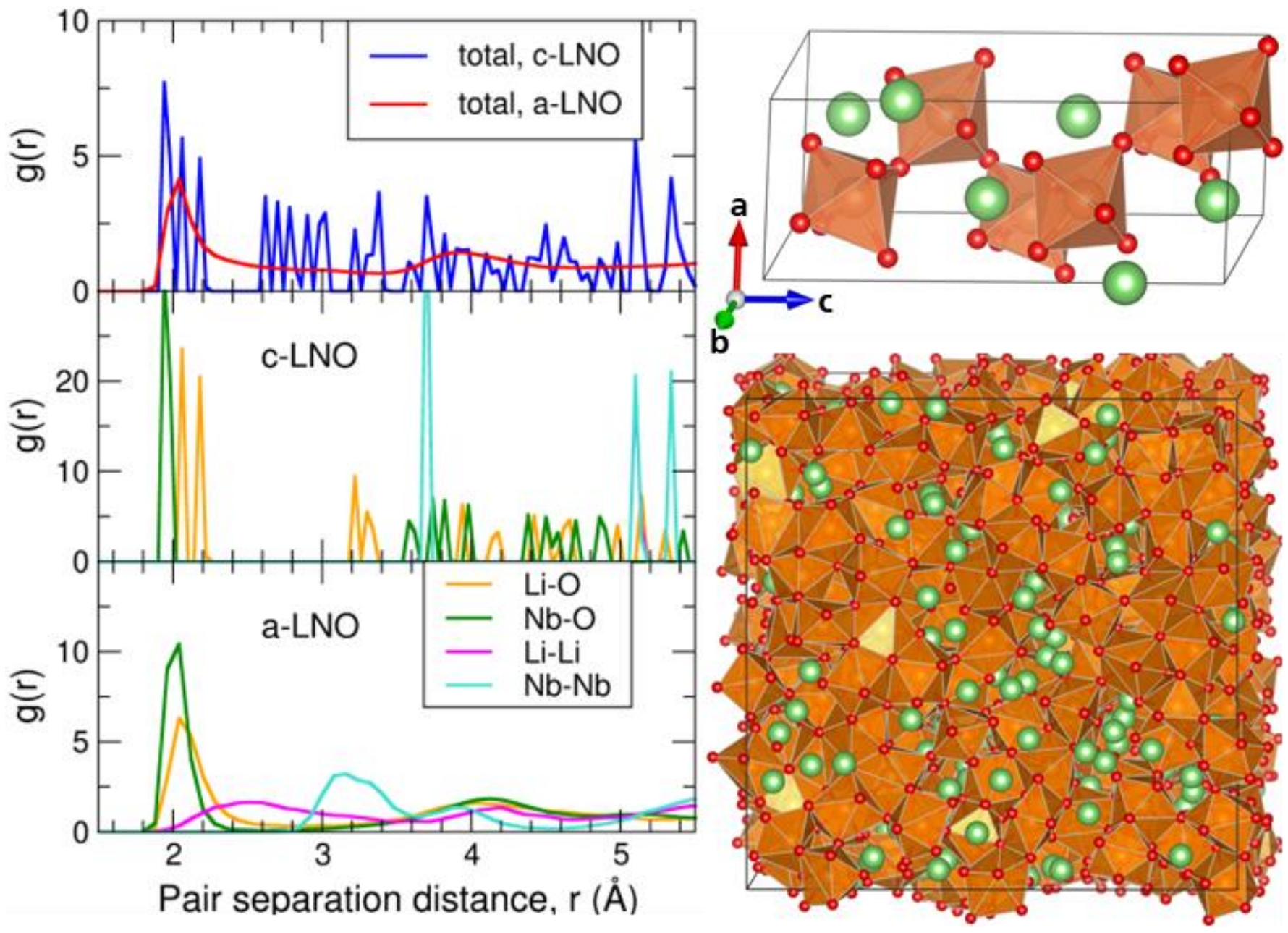

*Figure 2. Left: Total (top) and pairwise (center and bottom) RDFs of c-LNO and of a-LNO (average over 10 samples), in both cases for a bin size of 0.04 Å. Note that in c-LNO, the Li-Li and Nb-Nb peaks are exactly on top of each other. Right: conventional trigonal cell of c-LNO (top) and cubic box of a-LNO (bottom) with Li atoms in green, O atoms in red, and O-coordinated polyhedra around Nb atoms in orange.*

## 3. Results

### 3.1. Diffusion of Li ions in c-LZO

We set up a supercell with the dimensions of 5×3×5 conventional c-LZO cells specified in Table 1, corresponding to the stoichiometry $Li_{600}Zr_{300}O_{900}$. An initial heating simulation of this system gave a melting point of about 2500 K. To ensure that the structure stays stable during the calculation of the MSDs, a temperature range considerably below this value, namely 1400 – 2025 K, was chosen for the MD simulations. Two sets of simulations were performed. In the first set, we removed one Li ion from the system to create a Li vacancy. The diverging energy of a charged periodic cell is treated in the standard way by ignoring the term for $q = 0$ in the reciprocal space part of the Ewald sum (which corresponds to applying a neutralizing background charge) [57]. In the second set of simulations, the stoichiometry was not changed. For both sets, we performed 10 runs at each considered temperature. Those 10 systems differ only by the seed for the random number generator used for the Gaussian profile of initial velocities corresponding to 1 K. After heating the systems from there to the temperatures of choice via isotropic NPT runs, the MSDs, denoted as $m_k(\tau)$ in the following, were obtained for the Li, Zr, and O ions in subsequent NVT runs from their trajectories $\vec{r}_i(\tau)$, according to:

$$m_k(\tau) = \frac{1}{N_k}\sum_{i=1}^{N_k}\left(\vec{r}_i^{\,k}(\tau) - \vec{r}_i^{\,k}(0)\right)^2. \quad (1)$$

The index $k$ distinguishes Li, Zr, and O, and $N_k$ denotes the number of atoms of type $k$ in the system. Simulations were performed for a maximum time span of $\tau_{\max} = 50$ ns.

The MSDs of Li are shown in Figure 3 for the systems with and without a Li vacancy at different temperatures. In the case that a vacancy is present, it mediates the diffusion by successive jumps of neighboring Li ions into the vacant sites, which leads to the rather smooth, mostly linearly increasing lines at the lower temperatures as expected by the theory (Einstein-Smoluchowski relation) [58]. Occasionally, there are additional steps of the MSD, which are more frequent at the highest temperature shown for this set, 1625 K. In the systems without vacancy, one would not expect diffusion of Li in a stable crystal structure, however, the MSDs increase linearly, as well. When looking at the individual curves in detail, i.e., at smaller time intervals, one can identify that the ascents are completely due to steps, which happen more frequently at the higher temperatures considered for this set. In between those steps, the curves are flat, except for thermal oscillations. As will be discussed in detail in Section 4, this behavior originates from collective Li-Li site exchange processes. The other ions in the system, Zr and O, do not show an increase of the MSD of this kind in any of the sets, which confirms the structural stability of c-LZO during the simulations. In the systems with a Li vacancy, there are very rare events of a Zr ion jumping into the vacancy, but oxygen is never found there, as this is energetically highly unfavorable.

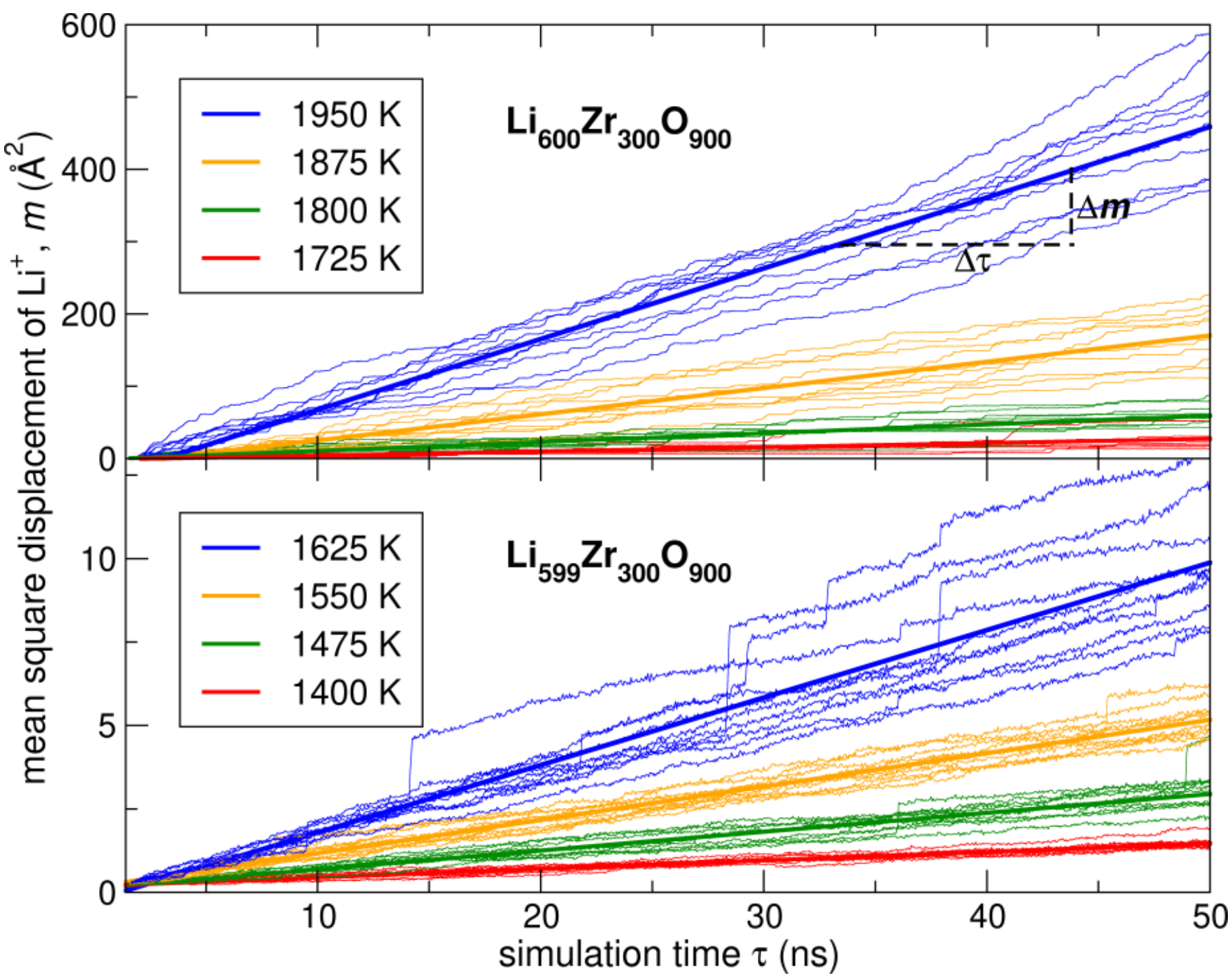

*Figure 3. MSDs of Li ions during MD simulations at different temperatures for systems with (bottom) and without (top) Li-vacancy. The thick lines are averages over linear regression lines of each of the 10 MSD gradients at each temperature.*

In both sets, for each of the 10 considered systems at each temperature, a linear regression of $m_{\mathrm{Li}}(\tau)$ was performed in the range from $\tau = 0$ to $\tau = \tau_{\max}$ and averaged over the systems, leading to a mean value of $\Delta m/\Delta\tau$ for each case (as indicated in Figure 3). From the expressions of the diffusivity, on the one hand as the proportionality constant between MSD and time ($D = m/6\tau$), and on the other hand as an exponential function of an effective activation energy $\Delta E$ of the underlying atomic migration processes ($D \propto \exp(-\Delta E/k_{\mathrm{B}}T)$), one obtains [58]:

$$\ln D = \ln\left(\frac{\Delta m}{6\Delta\tau}\right) = \ln\nu - \frac{\Delta E}{k_B T}, \quad (2)$$

with the preexponential factor $\nu$ and Boltzmann's constant $k_B$. As shown in Figure 4, plotting $D$, as calculated from the averaged values of $\Delta m/\Delta\tau$, on a logarithmic scale versus the inverse temperature (Arrhenius plot) leads to linear lines as expected from Equation (2).

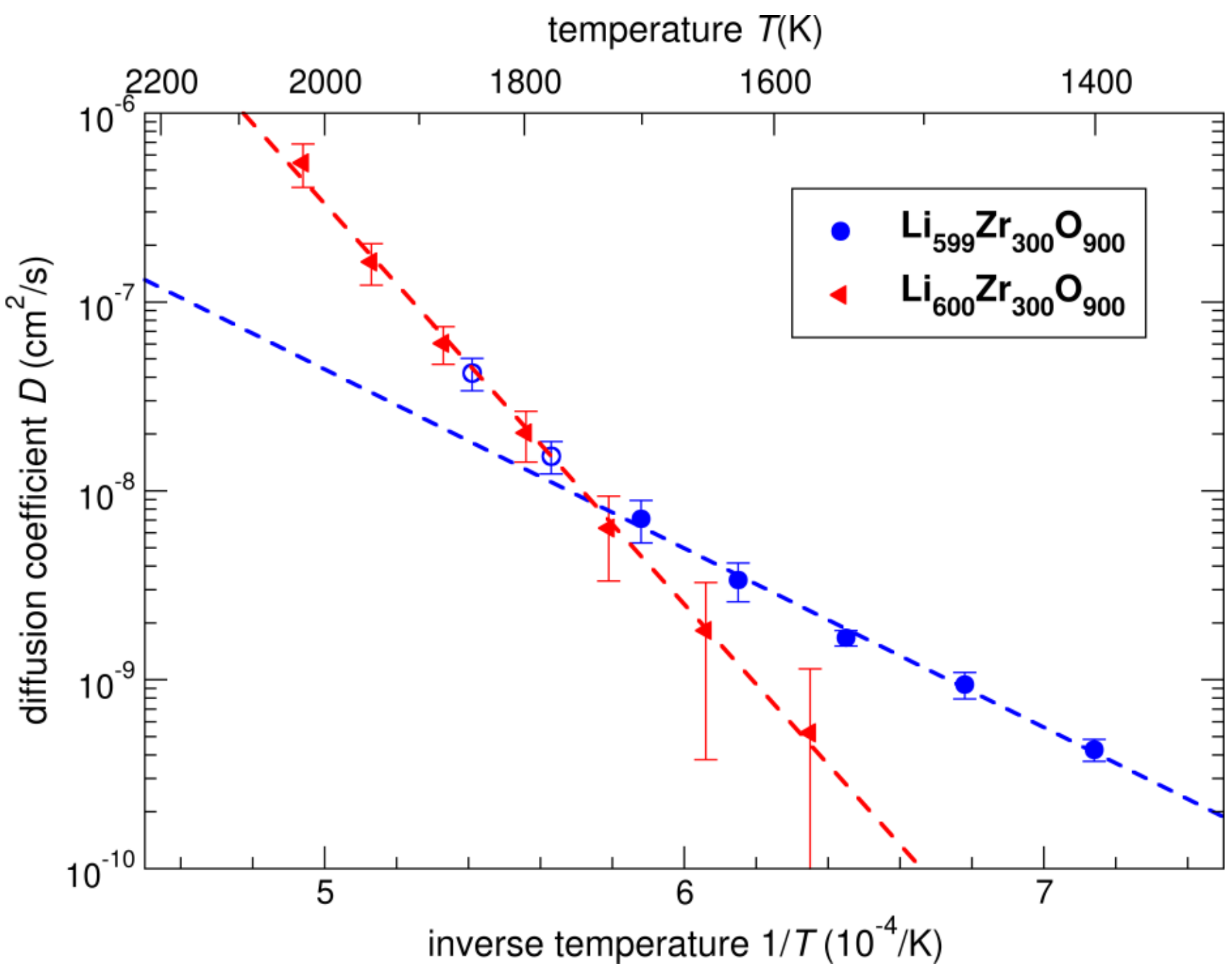

*Figure 4. Diffusivity of Li ions in c-LZO with and without Li vacancy, with linear regression lines through the data points which are represented by the respective filled symbols. The open blue circles were obtained for systems with vacancy. Error bars result from the standard deviations of the slopes $\Delta m/\Delta\tau$ determined from the 10 systems for each data point.*

One can clearly discriminate two different slopes. For the systems without vacancy, all the data points lie on one line, indicating that in these compounds, Li migration takes place via the same mechanism, with an effective activation energy $\Delta E = 4.2$ eV. At low temperatures, this energy barrier is too high to be overcome within the considered simulation time, and the large error bars of the data points belonging to $T = 1575$ K and 1650 K result from the occurrence of only a very low number of steps in the MSD curves. The vacancy-mediated process has an activation energy $\Delta E = 2.2$ eV. This makes it more likely to take place at the lower temperatures, which leads to the higher values of $D$. The occasional vertical steps in the MSD curves of the systems with vacancy are too rare to dominate the Li diffusion behavior up to about 1700 K. This however changes for higher temperatures, where the migration via vacancy-mediated jumps becomes overcompensated by the site-exchange mechanism, as seen by the open blue symbols in Figure 4.

A similar analysis of the Li diffusion in c-LNO was not possible. In the MD simulations, the structure became unstable with respect to the liquid phase at about 1500 K, which agrees well with the experimental melting point [59]. It turned out that this temperature is below the range of temperatures where in the MD simulations hopping processes could be observed during reasonably long time spans $\tau_{max}$ in a number sufficient for a statistical Arrhenius evaluation.

### 3.2. Diffusion of Li ions in a-LZO and a-LNO

For the MSD calculations of the amorphous phases, we chose the structures and system sizes as described in Section 2.2, i.e., each 10 systems for each composition of LZO and LNO. Additional vacant sites were not considered at first. To define the range of temperatures for the simulations, the systems were first heated up to 4000 K in isotropic NPT runs.

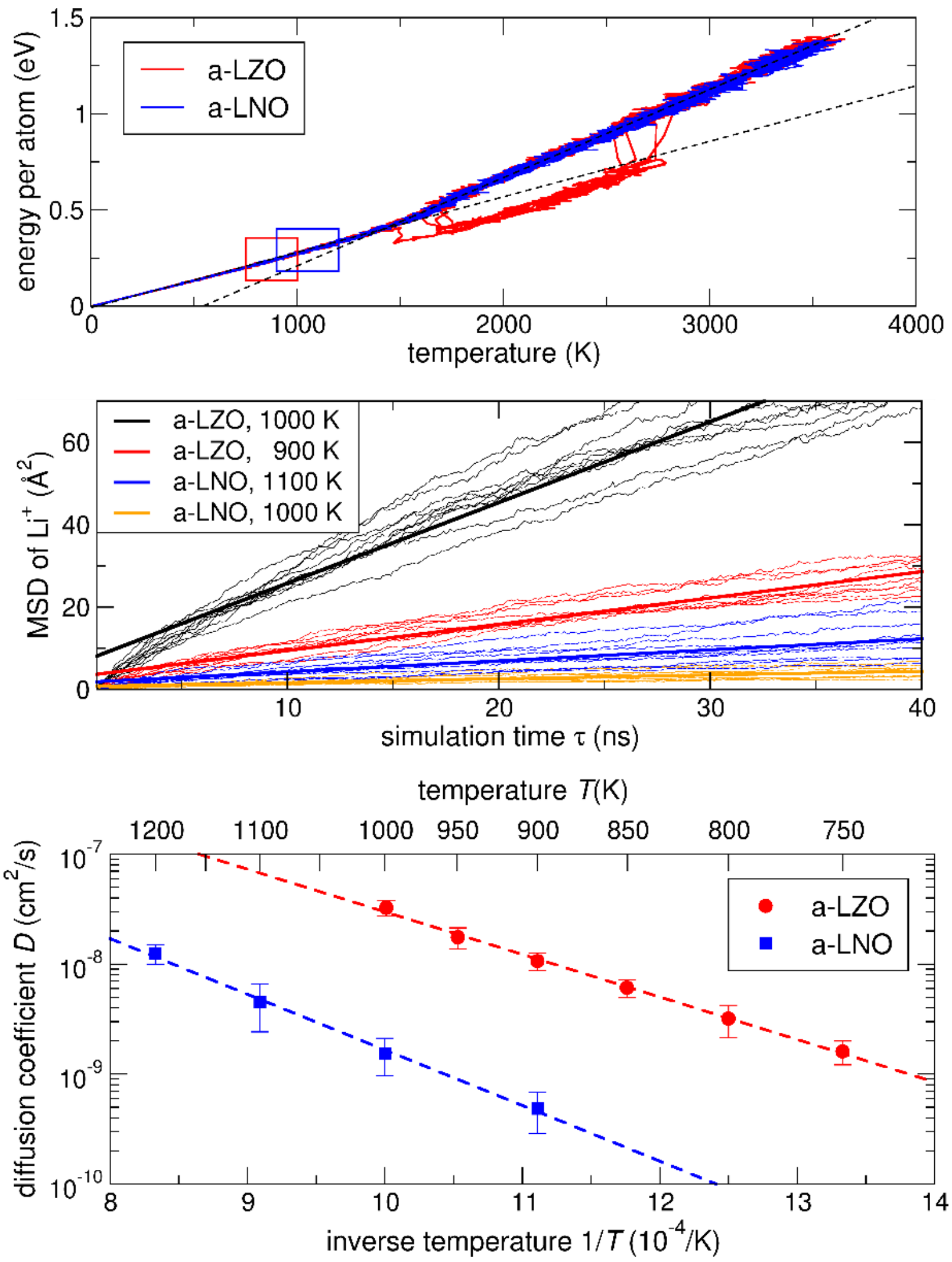

*Figure 5. Top: Energies (per atom) of 10 a-LZO and 10 a-LNO systems upon heating in an isotropic NPT simulation (relative to the respective energies at $T = 0$ K). Different slopes of the amorphous and liquid phases are indicated by dashed lines. The rectangles mark the temperature regions in which the MSD values were calculated. Center: MSDs of Li ions from NVT simulations in a-LZO and a-LNO at an exemplary set of temperatures, together with thicker lines representing averaged regression lines of the individual MSD curves. Bottom: Diffusion coefficients of a-LZO and a-LNO in an Arrhenius plot. Error bars result from the spreads of the 10 MSD curves for each temperature.*

As shown in Figure 5 (top), the curves of energy per atom versus temperature increase linearly until about 1500 K. Above this temperature, structural instabilities arise: all ten a-LNO systems transform into the liquid phase characterized by a different slope of the curve of energy versus temperature. A part of the a-LZO systems behaves in the same way, and a part shows recrystallization corresponding to a drop in energy, but all transform into the liquid phase at higher temperatures, too. To ensure that such transformations do not interfere with the MSD calculations, we chose the temperature ranges well below the transformation region: 750 K – 1000 K for a-LZO and 900 K – 1200 K for a-LNO. As for c-LZO, the MSDs of the different ionic species in the amorphous phases were recorded in NVT runs (here for $\tau_{\max} = 40$ ns) after heating the systems via isotropic NPT simulations to various temperatures in the specified ranges. Figure 5 (middle) shows an exemplary set of MSD curves. Comparable to the situation in c-LZO, the MSD curves increase, on average, linearly, with a limited spread between the 10 systems considered in each case. Pronounced steps as in the crystalline phase are not seen. The Arrhenius analysis according to Equation (2) based on the averaged slopes of the regression lines results in well separated linear curves for the two phases as depicted in Figure 5 (bottom). From those lines we extracted the activation energies of $\Delta E = 0.8$ eV for a-LZO and $\Delta E = 1.0$ eV for a-LNO. These similar values indicate the same underlying diffusion mechanism in both compounds. As we will discuss in Section 4, this mechanism can also be understood as collective Li-Li site exchange events as in stoichiometric c-LZO.

For that crystalline compound, as described in Section 3.1, the presence of a Li vacancy enables a different diffusion mechanism which is dominant at lower temperatures. Such a behavior was not observed for the amorphous phases. Even in case of considering samples with Li-substoichiometry of up to 10 % (by randomly removing Li ions from the systems), the averaged slopes of the MSD curves did not discernibly differ from those obtained for the unaltered compounds at the same temperatures. At temperatures below the regions specified in Figure 5 (top), there was no constant increase of the MSD curves in the substoichiometric systems within the simulation time. Hence, an effective activation energy of a diffusion process in the amorphous phase, where Li ions successively move to areas of free volume resulting from the substoichiometry (in analogy to the vacancy-mediated diffusion in crystalline phases) must have a higher activation energy value than the site exchange process.

## 4. Discussion

As described in Section 3, the MD simulations of c-LZO at high temperature revealed a mobility of Li ions higher than that of the vacancy-mediated hopping by maintaining the crystal structure. This can be inferred on the one hand from the observation that neither of the other ionic species (Zr or O) shows such an increase of the MSD except for the amplitude of thermal vibrations, and on the other hand from the result that the averaged positions of Li ions are unchanged even though the MSD increases steadily. Repeated sudden events of Li-ion movement within the structure cause this behavior, which corresponds to the vertical steps in the MSD curves, and the lower the temperature, the less frequent those events occur. To analyze the underlying mechanism in detail, we selected a single run of stoichiometric c-LZO at a temperature of 1650 K with only a few, well separated and clearly distinguishable MSD steps during the simulation (see Figure 6 (top)). Both steps taking place in this run occur within about 0.15 ns (150 ps), and they are the sum of multiple individual but concurrent ion migration events which last for only a fraction of this time span.

This is exemplarily shown in Figure 6 (mid) for three neighboring Li ions (denoted by A, B, and C), which were identified to be involved in the total migration event of the first MSD step. Within only 2 ps, those ions move away from their averaged equilibrium positions by about 3 Å, which is the Li-Li nearest neighbor separation (*cf.* Figure 1). Tracking the mutual distances between the involved ions (depicted in Figure 6 (bottom)) reveals pronounced dips, at most down to about 2 Å, in correlation with their migration behavior. Such a correlation was found as a common principle in the migration events of all the Li ions contributing to the MSD steps. For the exemplary ions, chronological snapshots of the configuration at representative points in the simulation time (Figure 6) elucidate the mechanism: neighboring Li ions occasionally form Li-Li pairs which can undergo rotation, thereby exchanging sites with each other. Since all sites are occupied, this is only possible as a collective event involving several partners. In the example, more than the depicted ions need to be involved in the whole process, because the position, where ion A finally settles, was previously occupied by another Li ion. Comparing the coordinates of all Li ions before and after the complete MSD step yields that 55 of the 600 Li ions moved to nearest neighbor sites and 16 to even larger distances, which altogether lead to the MSD increment of about 2 Å$^2$. A migration process mediated by interstitial Li ions and the successive hopping of Li-Li pairs was also reported for solid electrolyte compounds crystallizing in the NASICON framework, such as LATP ($Li_{1+x}Al_xTi_{2-x}(PO_4)_3$) [60], [61]. However, for that process to happen, additional Li ions occupying interstitial positions must be present in the structure, and, unlike the site-exchange process described above, no more than three Li ions, two forming the pair and one on a neighboring site, are involved in a single migration event.

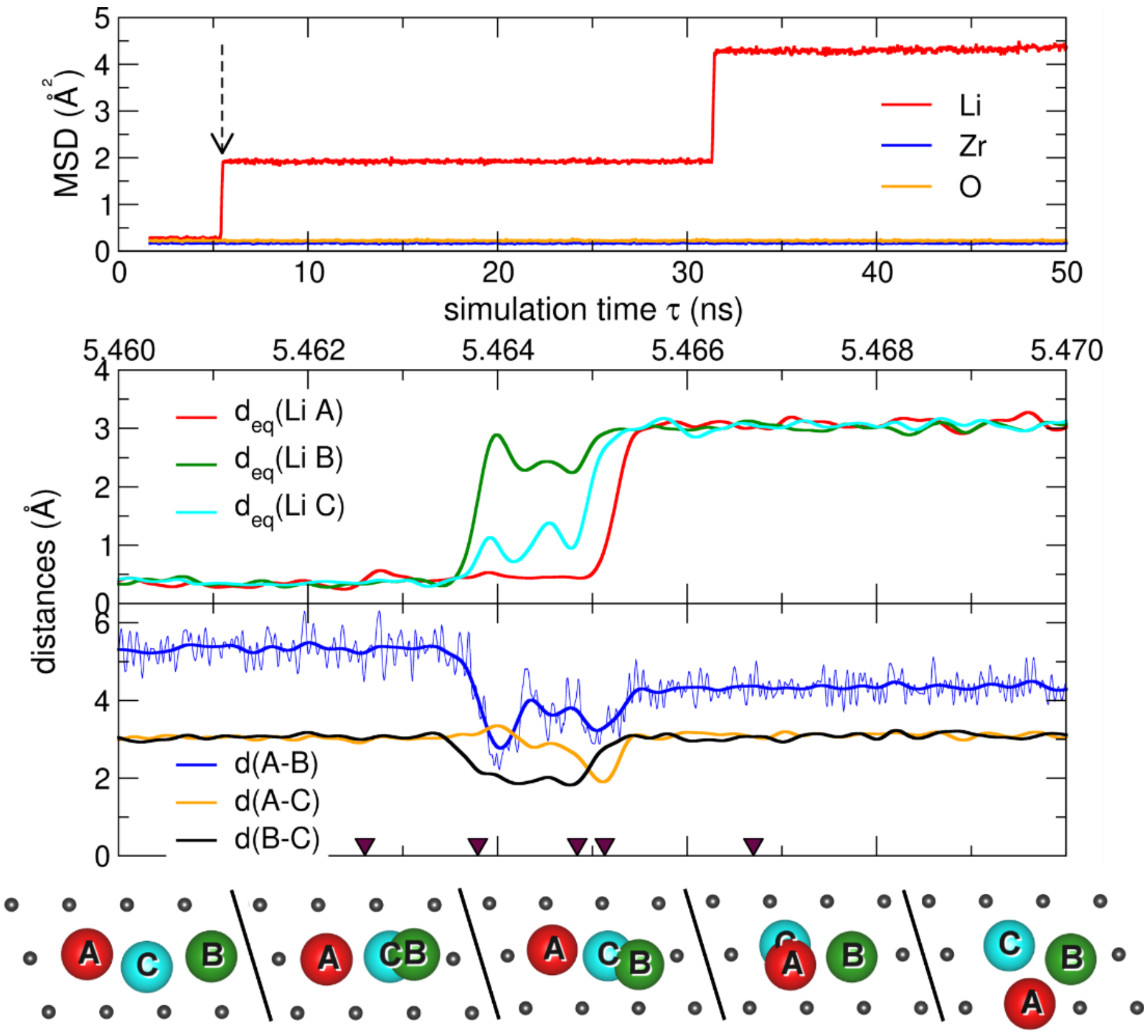

*Figure 6. Depiction of the site-exchange mechanism in c-LZO. Top graph: MSD of Li, Zr, and O ions in one run at $T = 1650$ K. The arrow indicates where the atomic distances are analyzed further. Central graph: distances $d_{eq}$ of three exemplary Li ions [denoted as A, B, C] from their equilibrium positions before the migration (averaged over thermal vibrations). Bottom graph: mutual distances $d$ between the Li ions A, B, and C. For better visibility, the curves were smoothed by Gaussian smearing as shown exemplarily for $d$(A-B). This was also done for the curves of $d_{eq}$ in the central graph. The five black triangles on the x-axis of the bottom graph indicate the instants in the simulation time at which the two-dimensional snapshots of the configuration from left to right are depicted. Note that Li ions A, B, and C do not necessarily all lie in this plane. Little black spheres indicate the sublattice of the Li ions.*

In the stoichiometric phases treated in this work, the occurrence of Li-Li pairs enabling the Li mobility in the way illustrated in Figure 6 is intimately related to the formation of Frenkel defects. By building a defect complex of a vacancy and an interstitial, the latter necessarily approaches other ions of the same kind in the environment. The observation that none of the other two ionic species show a similar self-diffusion process in the structure is consistent with the Li Frenkel defect having the lowest formation energy compared to those involving Zr or O. This energetic hierarchy of Frenkel defects in c-LZO was derived both by DFT-based [62] and potential-based [63] studies. With 5.77 eV, the DFT-derived defect formation energy of the Li Frenkel defect is rather high [62], and accordingly, a self-diffusion mechanism as the one described above is unlikely to occur in this material up to high temperatures.

Using NMR, Bottke et al. observed an extremely slow Li diffusion in polycrystalline c-LZO and describe it as resulting from an exchange of two Li ions between neighboring, symmetrically inequivalent sites [64]. A comparison of the diffusivity given in that paper ($10^{-22}$ $m^2/s$ at 310 K) with ionic conductivity data reported by Hellstrom and van Gool [17] leads to an activation energy of 0.94 eV for the Li-ion diffusion process in c-LZO described by Bottke et al. [64]. This activation energy is considerably lower than both, the value derived from DFT results (2 eV [18]) and from simulation results using the Morse potential in this work (4.2 eV) for a pure site-exchange process in a perfect LZO crystal. An explanation may be a different diffusion mechanism in the experimental studies, facilitated, e.g., by the presence of extended structural defects (lattice dislocations or grain boundaries) in the polycrystalline samples. For vacancy mediated diffusion, DFT-NEB calculations report activation energies of 0.75 eV [18], 0.55 eV [63], and 0.48 eV [65] for the hopping of a Li ion into a neighboring vacant site, and Baklanova et al. measured $\Delta E = 0.5$ eV [19]. With $\Delta E = 2.2$ eV, the value derived from the MD simulations in this work for the vacancy-mediated process is considerably higher. It is well known that the BV method, based on which the Morse potential applied in this work is constructed, overestimates the activation energies of ionic migration processes [66]. As described in Section 2.2, this potential predicts the edge lengths of the cell and the monoclinic angle to be shorter in the range of 1.6 – 7.0 % than the experimental lattice parameters and the values obtained by DFT calculations [19]. In a denser structure, energy barriers of migration processes must obviously be higher since equally charged ions are closer to each other. This can explain the deviations between DFT-derived activation energies and those resulting from the employed Morse potential.

Qualitatively consistent with experimental studies is the general result that the diffusion coefficient is higher by many orders of magnitude in an amorphous phase than in its crystalline counterpart, as, e.g., reported for Li ion diffusion in Li niobate glasses [23], [25], [26], [67], [68], for oxygen diffusion in a-$Al_2O_3$ [69], or for Si diffusion in a-Si-Ge systems [30]. For a-LZO, we obtained $\Delta E = 0.8$ eV from the Arrhenius analysis in Section 3.2. As described there, a substoichiometry of Li did not change the diffusion behavior in the considered temperature range. To further analyze the mechanism in the amorphous phase, we chose one exemplary run at $T = 900$ K and cooled the system down to 1 K once before and once after recording the MSDs of the different species over the simulation time $\tau_{\max}$ [as depicted for Li in Figure 5 (mid graph)]. An alternative calculation of the MSDs by just comparing the positions of the ionic sites of those two structures lead to values of 0.6 $\text{Å}^2$ for Li, 0.3 $\text{Å}^2$ for Zr, and 0.4 $\text{Å}^2$ for O. This indicates that the structure does not considerably change during the simulation at 900 K. However, the MSDs obtained when following each ion individually were 24.6 $\text{Å}^2$ for Li, 0.4 $\text{Å}^2$ for Zr, and 1.6 $\text{Å}^2$ for O. While the MSDs calculated for Zr in the two different ways are almost identical, the values for Li differ strongly. This can only be explained reasonably by Li ions undertaking multiple site exchange processes during the simulation by keeping the whole structure intact. Zr ions mainly stay at their positions and O ions migrate only very rarely. In c-LZO, the site exchange was realized by the formation and subsequent rotation of Li-Li pairs with reduced distance compared to the equilibrium distance in the crystal structure. As can be seen in the partial RDFs of c-LZO and a-LZO in Figure 1 (left graph), the Li-Li peak, although strongly broadened, is significantly shifted to the left in the amorphous phase. The pair formation process as the necessary precursor for a site exchange may therefore occur more easily, corresponding to a lower activation energy for Li migration in a-LZO.

Due to the strong electrostatic attraction between $Zr^{4+}$ and $O^{2-}$ ions, the configuration of a-LZO is mainly defined by connected O polyhedra around the Zr ions as shown in Figure 1, with Zr-O distances and, concomitantly, Zr-Zr distances between the polyhedron centers close to the respective equilibrium values in the crystal phase. Li ions are randomly arranged on the interstitial sites in this network, constrained by also maintaining the favorable Li-O distance of the crystal structure. Since the absolute strength of the electrostatic Li-Li interaction is the lowest among all the ionic pairs, keeping those ions at a large distance is least respected during the optimization of the configuration in the rapid quenching when the amorphous phase is formed. This can explain the reduced Li-Li separation in a-LZO compared to c-LZO. A similar explanation holds for the formation of the niobium-containing system, a-LNO, where, as in a-LZO, the Li-Li peak is located at a lower value as in the crystal phase (see Figure 2). With the center of mass at about 2.5 Å, it is above the position of the Li-Li peak in a-LZO of about 2.3 Å. Due to this difference, the Li ion migration via site exchange, which was detected in a-LNO, too, has a higher activation energy (1.0 eV) as in a-LZO (0.8 eV).

While, as described above, the discrepancy in activation energies obtained by the procedure in this work and activation energies reported from experiments or DFT calculations is considerable for c-LZO, the activation energies of the amorphous phases are overestimated much less by the model potential approach. For Li diffusion in a-LNO, experimental values between 0.4 eV and 0.83 eV were obtained [23]–[26], [70]. Using isotope heterostructures, Rahn et al. measured $\Delta E = 0.7$ eV for Li-poor a-LNO (compared to stoichiometric $LiNbO_3$) [25], which is in reasonable agreement to $\Delta E = 1.0$ eV obtained by our simulations. However, the similarity within error limits of the preexponential factors of Li diffusion in a-LNO reported in Ref. [25] and in c-LNO in a study by the same authors [21] lead these authors to the conclusion of a similar diffusion mechanism, namely a vacancy-mediated one in the crystalline phases, and via vacancy-like defects in the amorphous phases, where a Li atom is locally missing in the short range order. While this process would differ from the site-exchange mechanism observed in the simulations described in this work, Rahn et al. later reported the Li ion diffusivity also in Li-rich a-LNO (compared to stoichiometric $LiNbO_3$) [26], where the concept of such vacancy-like defects is less obvious. The reported activation energy of 0.83 eV in the Li-rich amorphous phases is even closer to the value obtained by our simulations (1.0 eV). An extrapolation of the corresponding Arrhenius-plot line displayed in Figure 5 (bottom panel, blue line) to $1/T = 0$ yields a preexponential factor of $1.9 \cdot 10^{-8}$ $m^2/s$ which is lower than the values reported by Rahn et al. for Li-poor ($5 \cdot 10^{-7}$ $m^2/s$) [25] and Li-rich ($1.2 \cdot 10^{-5}$ $m^2/s$) amorphous phases [26]. Keeping in mind that those values were also obtained from extrapolations over a large range of diffusivities versus inverse temperatures, where small changes in activation energies amount to deviations of several orders of magnitude in the preexponential factor, the agreement between their measured and our calculated values is acceptable. Therefore, it is reasonable to assume that the measured Li diffusivities in the a-LNO samples can be explained to some degree of confidence by a site-exchange mechanism as described in this work. To the best of our knowledge, there are no reports of activation energies for a-LZO, but Kuwata et al. measured the Li-ion conductivity in several amorphous Li-containing oxides and extracted activation energies in the range of 0.6 – 0.7 eV [71], which is close to the value we determined for a-LZO.

## 5. Summary and Conclusion

We applied classical MD simulations with a Morse potential model based on the Bond Valence method, which is widely used to analyze static energy landscapes of ionic migration, to study the dynamic diffusion behavior of Li ions in crystalline and amorphous Li-Zr-O and in amorphous Li-Nb-O systems. First, the equilibrium crystal structures of $Li_2ZrO_3$ and $LiNbO_3$ were checked and confirmed to be stable by the potential, although the lattice parameters are underestimated. The amorphous phases, obtained by quenching equilibrated liquid phases, are composed of networks of distorted metal-oxygen polyhedra with Li ions distributed interstitially in the amorphous Zr-O and Nb-O host networks. Steps were found in the mean square displacement curves of Li for both, crystalline and amorphous systems; for the latter the steps occurred much more frequently and at lower temperatures. The origin of those steps is a collective site-exchange process of Li ions, related to the formation of Frenkel pair defects and subsequent rotation of Li-Li pairs. Smaller distances between Li ions in the amorphous phases, compared to their separation in the crystal structures, enable the site exchange mechanism with much lower activation energies. Altogether, such large drops in activation energies of ionic diffusion when going from a crystalline phase to an amorphous phase are consistent with widely reported findings in the literature for diverse materials. For the systems studied in this work, the activation energies are quantitatively overestimated by the employed Morse potential with respect to experimental values. However, the outlined methodology and the discussed results shed light on a hitherto rarely described diffusion mechanism in ionic metal-oxide materials relevant for ion-battery applications.

**Declaration of Competing Interest**

None.

**Acknowledgments**

This work was funded by TotalEnergies OneTech (TE). The authors thank Christoph Sachs (TE), Jamie Greer (Air Liquide Advanced Materials), Vincent Pele and Christian Jordy (Saft) for fruitful discussions. The work was enabled in-part by TE's HPC for R&D infrastructure and services in Houston, Texas, using the Cedar Cluster. Atomistic structure models are visualized using VESTA [72].